\documentclass[12pt]{article}
\usepackage{geometry}                
\usepackage{graphicx}
\usepackage{bm}
\usepackage{amssymb,amsmath,amsthm}
\usepackage{mathrsfs} 
\usepackage[applemac]{inputenc} 
\def\TODAY{24 January 2013; 26 June 2013}
\title{\vskip-2cm \bf Bounds on variable-length \\ compound jumps}
\author{{\Large Petarpa Boonserm}\\[5pt]
Department of Mathematics and Computer Science\\ 
Faculty of Science, Chulalongkorn University\\
Phayathai Road, Pathumwan\\
Bangkok 10330, Thailand\\[5pt]
{\sf \small petarpa.boonserm@gmail.com} \\[7pt]
{\Large Matt Visser}\\[5pt]
School of Mathematics, Statistics, and Operations Research\\
Victoria University of Wellington \\
PO Box 600, Wellington 6140, New Zealand\\[5pt]
{\sf \small matt.visser@msor.vuw.ac.nz}  }
\date{\TODAY;  \LaTeX-ed  \today}                                           
\begin{document}
\maketitle
\begin{abstract}
{\small\noindent
In Euclidean space there is a trivial upper bound on the maximum length of a compound ``walk'' built up of variable-length jumps, and a considerably less trivial lower bound on its \emph{minimum} length.  The existence of this non-trivial lower bound is intimately connected to the triangle inequalities, and the more general ``polygon inequalities''.
Moving beyond Euclidean space, when a modified version of these bounds is applied in ``rapidity space'' they provide upper and lower bounds on the relativistic composition of velocities. 
Similarly, when applied to ``transfer matrices'' these bounds place constraints either (in a scattering context) on transmission and reflection coefficients,  or (in a parametric excitation context) on particle production. 
Physically these are very different contexts, but mathematically there are intimate relations between these superficially very distinct systems. 
Journal of Mathematical Physics {\bf54} (2013) 092105 [arXiv:1301.7524].\\
DOI: 10.1063/1.4820146 
}
\enlargethispage{30pt}
\end{abstract}

\def\d{{\mathrm{d}}}
\newcommand{\scri}{\mathscr{I}}
\newcommand{\sun}{\ensuremath{\odot}}
\def\J{{\mathscr{J}}}
\def\sech{{\mathrm{sech}}}
\def\n{\mathbf{n}}
\def\x{\mathbf{x}}
\def\R{{\mathbb{R}}}
\def\C{{\mathbb{C}}}
\clearpage
\hrule
\tableofcontents
\bigskip
\hrule
\clearpage

\section{Background}

One is often confronted with physical or mathematical situations where some complicated process can be built up by compounding (that is, chaining together) a number of simpler but not necessarily equal individual steps. Examples (by no means an exhaustive list) include compounding a series of variable-length jumps in physical space,  the relativistic composition of multiple velocities, and the composition of transfer matrices for scattering from multiple distinct (non-overlapping) barriers. 

An interesting and pragmatically useful question is whether information concerning the individual steps can be used to place useful bounds on the overall compound process. Herein, we present examples of several such phenomena.
From a purely technical perspective, this discussion is largely based on the analysis of compound scattering processes presented in reference~\cite{Boonserm:2011}, but the applications will be completely different:
\begin{enumerate}
\item  
There is a simplification of the upper and lower bounds of that article to variable-length compound jumps in ordinary Euclidean physical space.
\item
There is a modification of the upper and lower bounds of that article to the special relativistic composition of velocities.
\end{enumerate}

Mathematically, the intimate relationship between special relativity and quantum scattering is due to the fact that the Lorentz group and group of transfer matrices are both Lie groups, with closely related though not identical Lie algebras. Specifically, the Lorentz group can be represented by $SO(3,1)$, which is locally isomorphic to $SL(2,\C)$, whereas the set of transfer matrices form a representation of $SU(1,1)$, which is locally isomorphic to $SL(2,\R)$. See, for example, the recent review article~\cite{geometric} and references therein. 
(For other relevant background material see for instance~\cite{moller1952, jackson1998, O'Donnell:2011, stapp1956, fisher1972, ferraro1999} on composition of velocities in special relativity and~\cite{Merzbacher, Mathews, Singh, Sanchez-Soto, pedagogical, Peres, Kowalski, Korasani, SS1, SS2} on quantum scattering.) 

It is the structural similarity between the Lie algebras of $SL(2,\C)$ and $SL(2,\R)$, and the relation between velocities and rapidities, versus the relation between transmission probabilities and Bogoliubov coefficients,   that underlies the close mathematical similarities between relativistic composition of velocities and the compounding of transfer matrices. 
For instance, an arbitrary boost can always, up to a 3-dimensional rotation $R$, be written as
\begin{equation}
{\cal B}  =  R \; \exp\left( \xi 
\left[ \begin{array}{cccc}0&1&0&0\\1&0&0&0\\0&0&0&0\\0&0&0&0\end{array}\right] 
\right) \; R^{-1},
\end{equation}
with the speed being related to the rapidity by $v = \tanh \xi$. In counterpoint, an arbitrary transfer matrix can always be written in the form~\cite{Boonserm:2011}
\begin{equation}
{\cal{T}} = \left[ \begin{array}{cc}\alpha&\beta^*\\\beta&\alpha^*\end{array}\right] = 
\left[ \begin{array}{cc}\cosh\Theta \;e^{i\phi}&\sinh\Theta \; e^{-i\psi} \\ 
\sinh\Theta\;e^{i\psi}&\cosh\Theta \;e^{-i\phi}\end{array}\right].
\end{equation}
It is then easy to see that
\begin{equation}
{\cal{T}} =
\left[ \begin{array}{cc}e^{i(\phi-\psi)/2}&0\\ 0&e^{-i(\phi-\psi)/2}\end{array}\right] 
\exp\left( \Theta \left[ \begin{array}{cc}0&1 \\ 
1&0\end{array}\right] \right)
\left[ \begin{array}{cc}e^{i(\phi+\psi)/2}&0\\ 0&e^{-i(\phi+\psi)/2}\end{array}\right],
\end{equation}
with the reflection probability being given by $\sqrt{R} = |r| = \tanh \Theta$. (See, for instance~\cite{Boonserm:2011}.)
Furthermore the appropriate subspaces of the Lie algebras of both of these Lie groups can be mapped homeomorphically (and even monotonically) to the Euclidean translations, which ultimately underlies the close connection to compound jumps in ordinary Euclidean space. Indeed, working with the Euclidean space formulation in some sense ``trivializes'' the bounds and makes clear the close connection between the lower bound and the triangle inequalities (or more generally the polygon inequalities).

\clearpage
\section{Variable length random walks in physical space}

Suppose we have a compound ``walk'' in physical where the individual step sizes (``jumps'') are fixed but variable, $\ell_1$, $\ell_2$, $\ell_3$, \dots,  $\ell_n$, but the directions $\n_i$ are arbitrary. What if anything can we say about upper and lower bounds on the net displacement
\begin{equation}
\x_{12\cdots n} = \sum_{i=i}^n \n_i \,\ell_i   ?
\end{equation}
Consider the two step case
\begin{equation}
\x_{12} = \n_1 \,\ell_1  + \n_2 \,\ell_2,
\end{equation}
then it is elementary that
\begin{equation}
|\ell_1-\ell_2| \leq |\x_{12}| \leq  \ell_1  + \ell_2.
\end{equation}
Furthermore it is also clear that for $n$ steps
\begin{equation}
|\x_{12\cdots n}| \leq   M_{12\cdots n} \equiv \sum_{i=1}^n\ell_i.
\end{equation}
But can one place a \emph{lower bound} on $|\x_{12\cdots n}| $? 
Yes, by a straightforward modification (and simplification) of the analysis of reference~\cite{Boonserm:2011}, for a three-step walk we assert (and shall soon prove):
\begin{equation}
|\x_{123}| \geq  \max\{ \ell_1 - \ell_2-\ell_3, \; \; \ell_2 - \ell_3-\ell_1, \;\; \ell_3 - \ell_1-\ell_2, \;\; 0\}.
\end{equation}
More generally, for an $n$-step walk we assert (and shall soon prove)
\begin{equation}
|\x_{12\cdots n}| \geq   \max\left\{  \ell_i - \sum_{j\neq i} \ell_j,  \;\; 0 \right\},
\end{equation}
or equivalently
\begin{equation}
|\x_{12\cdots n}| \geq  \max\left\{  2 \ell_i - \sum_{j=1}^n \ell_j,  \;\; 0 \right\}.
\end{equation}
We can also write this as
\begin{equation}
|\x_{12\cdots n}| \geq  m_{12\cdots n}  \equiv \max\left\{  2 \ell_i -  M_{12\cdots n}, \;\; 0 \right\}.
\end{equation}
(So, as is reasonably common notation,  we use $M$ to denote the maximum, and $m$ to denote the minimum.)

\section{Triangle and polygon inequalities}

To first see \emph{why} these lower bounds have any hope of working, it is useful to consider the triangle inequalities.

\subsection{3 steps}

A key observation is this:
\emph{The 3-step lower bound is non-trivial if and only if the three step-lengths, $\ell_1$, $\ell_2$, and $\ell_3$, violate the triangle inequalities.}
To see this, recall that for a three-step compound walk in physical space we asserted:
\begin{equation}
|\x_{123}| \geq  \max\left\{  \ell_1 - \ell_2 - \ell_3,   \ell_2 - \ell_3 - \ell_1,  \ell_3 - \ell_1 - \ell_2,   0 \right\}.
\end{equation}
Why this odd combination? This is related to the triangle inequalities in a quite elementary manner.
If $\ell_1$, $\ell_2$, and $\ell_3$ are the lengths of the sides of a physical triangle in Euclidean space then they must satisfy the triangle inequalities: The length of any one side of the triangle must be less than or equal to the sum of the lengths of the other two sides. 
That is:
\begin{equation}   
  \ell_1 \leq \ell_2 + \ell_3; \qquad
      \ell_2 \leq \ell_3 + \ell_1; \qquad
      \ell_3 \leq \ell_1 + \ell_2.
      \end{equation}
This implies:           
\begin{equation}
      \ell_1 - \ell_2 - \ell_3 \leq0; \qquad
      \ell_2 - \ell_3 - \ell_1 \leq0; \qquad
      \ell_3 - \ell_1 - \ell_2 \leq0.
\end{equation}       
Therefore in this situation:    
\begin{equation}
       \max\{  \ell_1 - \ell_2 - \ell_3,   \ell_2 - \ell_3 - \ell_1,  \ell_3 - \ell_1 - \ell_2,   0\}  = 0.
\end{equation}
That is, \emph{if} the quantities $\ell_1$, $\ell_2$, and $\ell_3$ are the lengths of the sides of a physical triangle in Euclidean space, \emph{then} 
there is no constraint on $|\x_{123}|$ apart from the trivial one: $ |\x_{123}| \geq 0$.     
Therefore, the lower bound on $|\x_{123}|$ is \emph{non-trivial} if and only if $\ell_1$, $\ell_2$, and $\ell_3$  \emph{cannot} be interpreted as the lengths of the sides of a physical triangle in Euclidean space. Furthermore, if the triangle inequalities are violated, then the non-trivial lower bound specifies the extent to which the 3 edges of the “would-be triangle” \emph{fail to close}. 

\clearpage
\subsection{$n$ steps}

Generalizing the above observation:
\emph{For $n$ steps the lower bound is non-trivial if and only if the polygon inequalities are violated.}
To see this, observe that for an $n$-step random walk the lengths $\ell_i$ can be interpreted as the physical lengths of an $n$-sided polygon if and only if all $n$ polygon inequalities are satisfied:
\begin{equation}
 \forall i  \qquad  \ell_i \leq \sum_{j \neq i} \ell_j. 
\end{equation}
These polygon inequalities are the natural generalization of the triangle inequalities.  They can be built up iteratively by subdividing any polygon into triangles, and then applying the triangle inequalities step-by-step.
That is
\begin{equation}
 \forall i  \qquad \ell_i - \sum_{j \neq i} \ell_j  \leq 0.
\end{equation}
But then
  \begin{equation}
     \max \left\{  \ell_i - \sum_{j \neq i} \ell_j , 0 \right\} = 0. 
 \end{equation}    
So if the lengths $\ell_i$ can be interpreted as the physical lengths of an $n$-sided polygon then     
there is no constraint on $|\x_{12\cdots n}|$ apart from the trivial one: $|\x_{12\cdots n}| \geq 0$.     
Therefore, the lower bound on $|\x_{12\cdots n}|$ is \emph{non-trivial} if and only if the $\ell_i$  \emph{cannot} be interpreted as the lengths of the sides of a physical $n$-sided polygon in Euclidean space. Furthermore, if the polygon inequalities are violated, then the non-trivial lower bound specifies the extent to which the $n$ edges of the “would-be polygon” \emph{fail to close}.

These observations, though mathematically rather straightforward, and possibly even trivial, make it much clearer \emph{why} the lower bounds take the form they do,  why there is any realistic hope of obtaining any non-trivial lower bound, and also why there is no realistic hope of a lower bound more stringent than the one we have enunciated.

\clearpage
\section{Proof of the lower bound}

Start by defining the sums ($j\in\{1,2,3,\dots, n\}$)
\begin{equation}
M_{123\cdots j} = \sum_{i=1}^j \ell_i.
\end{equation}
Then it is elementary that 
\begin{equation}
|\x_{123\cdots j}| \leq M_{123\cdots j}
\end{equation}
for all $j\in\{1,2,3,\dots, n\}$.

\subsection{Iterative version of the lower bound}
 
Now take
\begin{equation}
m_1 = \ell_1,
\end{equation}
and, for $j\in\{1,2,3,\dots, n-1\}$, iteratively define the quantities $m_{123\cdots(j+1)}$ by
\begin{eqnarray}
m_{123\cdots(j+1)} &=&  (\ell_{j+1} - M_{123\cdots j}) \; H(\ell_{j+1} - M_{123\cdots j}) 
\nonumber\\
&& + (m_{123\cdots j}-\ell_{j+1}) \; H(m_{123\cdots j}-\ell_{j+1}),
\end{eqnarray}
where $H(\cdot)$ is the Heaviside step function. We can equivalently re-write this iterative definition as
\begin{equation}
m_{123\cdots(j+1)} =  \max\left\{ \ell_{j+1} - M_{123\cdots j}, m_{123\cdots j}-\ell_{j+1}, 0 \right\}.
\end{equation}

\paragraph{Theorem:} By iterating the 2-step bounds one has
\begin{equation}
\forall n: \qquad m_{123\cdots n}   \leq |\x_{12\cdots n}| \leq M_{123\cdots n}.
\end{equation}

\paragraph{Proof by induction:} 
When we iterate the definitions for $M_{123\cdots j}$ and $m_{123\cdots j}$, then the first two times we obtain
\begin{eqnarray}
M_1 &=& \ell_1; \qquad \qquad m_1 = \ell_1;
\\
M_{12} &=& \ell_1+\ell_2; \qquad m_{12} = |\ell_1-\ell_2|.
\end{eqnarray}
Thus the claimed theorem is certainly true for $n=2$. Now apply mathematical induction: Assume 
that at each stage the interval $[m_{123\cdots j},M_{123\cdots j}]$ characterizes the  highest possible and lowest possible values of $|\x_{12\cdots j}|$.  Applying the 2-step bound to the pair $|\x_{12\cdots j}|$ and $\ell_{j+1}$ leads trivially to $|\x_{12\cdots(j+1)}|$ being bounded from above by
\begin{equation}
M_{123\cdots (j+1)} = M_{123\cdots j} + \ell_{j+1},
\end{equation}
and less trivially to being bounded from below by
\begin{equation}
m_{123\cdots (j+1)}=  \max\left\{ \ell_{j+1} - M_{123\dots j}, m_{123\cdots j}-\ell_{j+1}, 0 \right\}.
\end{equation}
This completes the inductive step. That is:
\begin{equation}
|\x_{12\dots(j+1)}| \in [m_{123\cdots (j+1)}, M_{123\cdots (j+1)}],
\end{equation}
as claimed.
\hfill $\Box$

\bigskip
\noindent
However these bounds are currently  defined in a relatively messy iterative manner. Can this be usefully simplified? Can we make the bounds explicit?

\subsection{Symmetry properties for the lower bound}

When we iterate the definitions of $M_{123\cdots j}$ and $m_{123\cdots j}$ a third time we see
\begin{equation}
M_{123} = \ell_1+\ell_2+\ell_3; \qquad 
m_{123} = \max\{ \ell_3 - (\ell_1+\ell_2),  |\ell_1-\ell_2|-\ell_3, 0\}.
\end{equation}
We can further simplify this by rewriting $m_{123}$ as
\begin{equation}
m_{123} = \max\{  \ell_1-\ell_2-\ell_3, \;\ell_2-\ell_3-\ell_1, \; \ell_3 - \ell_1-\ell_2, \;0\}.
\end{equation}
Note that this form of $m_{123}$ is manifestly symmetric under arbitrary permutations of the labels $123$. 
One suspects that there is a good reason for this.  In fact there is.

\paragraph{Theorem:}  The quantity $m_{123\cdots j} (\ell_i)$ is a totally symmetric function of the $j$ parameters $\ell_i$, where $i\in\{1,2,3,\cdots,j\}$.

\paragraph{Proof:}   By inspection the result is true for $m_1$, $m_{12}$, and $m_{123}$.
But this argument now generalizes. In fact, the easiest way of completing the argument is to provide an explicit formula, which we shall do in the next section.

\subsection{Non-iterative formula for the lower bound}

\paragraph{Theorem:}
\begin{equation}
\forall n :  m_{123\cdots n} =  \max_{i\in\{1,2,\dots n\}}\{ 2 \ell_i - M_{123\cdots n}, 0\} = \max_{i\in\{1,2,\dots n\}}\left\{  \ell_i - \sum_{k=1,k\neq i}^n \ell_k, 0\right\}.
\end{equation}
\paragraph{Proof by induction:}
We have already seen that the iterative definition of $m_{123\cdots j}$ can be written as
\begin{equation}
m_{123\cdots (j+1)} = \max\{\ell_{j+1}-M_{123\cdots j}, m_{123\cdots j}-\ell_{j+1}, 0\},
\end{equation}
which we can also rewrite as
\begin{equation}
m_{123\cdots (j+1)} = \max\{2 \ell_{j+1}-M_{123\cdots (j+1)}, m_{123\cdots j}-\ell_{j+1}, 0\}.
\end{equation}
Now apply induction. The assertion of the theorem is certainly true for $n=1$ and $n=2$, and has even been explicitly verified for $n=3$. Now assume it holds up to some $j$, then
\begin{eqnarray}
m_{123\cdots (j+1)} &=& \max\{2 \ell_{j+1}-M_{123\cdots (j+1)}, m_{123\cdots j}-\ell_{j+1}, 0\} 
\nonumber
\\
 &=& \max\left\{2 \ell_{j+1}-M_{123\cdots (j+1)}, \max_{i\in\{1,2,\dots j\}}\{ 2 \ell_i - M_{123\cdots j}, 0\}  -\ell_{j+1}, 0\right\} 
 \nonumber
\\
 &=& \max\left\{2 \ell_{j+1}-M_{123\cdots (j+1)}, \max_{i\in\{1,2,\dots j\}}\{ 2 \ell_i - M_{123\cdots (j+1)}, 0\}  , 0\right\} 
 \nonumber
\\
 &=&\max_{i\in\{1,2,\dots j,(j+1)\}}\{ 2 \ell_i - M_{123\cdots (j+1)} , 0\}.
\end{eqnarray}
This proves the inductive step. Consequently
\begin{equation}
\forall n: \quad m_{123\cdots n} =  \max_{i\in\{1,2,\dots n\}}\{ 2 \ell_i - M_{123\cdots n}, 0\},
\end{equation}
as claimed.
\hfill $\Box$

\bigskip
\noindent
To simplify the formalism even further, let us now define
\begin{equation}
\ell_\mathrm{peak} = \max_{i\in\{1,2,\dots n\}} \ell_i.
\end{equation}
(We shall use the subscript ``peak'' for the maximum of the individual $\ell_i$'s;  the words ``max'' and ``min'' will be reserved for bounds on the $n$-fold composition of the $\ell_i$.)
Then we can simply write
\begin{equation}
\forall n: \quad m_{123\cdots n} =  \max \{ 2 \ell_\mathrm{peak} - M_{123\cdots n}, 0\}.
\end{equation}
This is perhaps the simplest way of presenting the lower bound.

\clearpage
\section{Relativistic composition of velocities}

Let us now apply the Euclidean space result derived above to a more subtle situation;  the relativistic composition of velocities.
(For general background see references~\cite{moller1952, jackson1998, O'Donnell:2011, stapp1956, fisher1972, ferraro1999}.)

\subsection{Collinear velocities}

When it comes to the relativistic composition of velocities the key thing is to note that for a pair of collinear (parallel or anti-parallel) velocities we have
\begin{equation}
v_{12} = {v_1+v_2\over 1+ v_1 v_2},
\end{equation}
which implies
\begin{equation}
 {\left| \; |v_1|-|v_2| \; \right| \over 1- |v_1| |v_2|}   \leq |v_{12}| \leq {|v_1|+|v_2|\over 1+ |v_1| |v_2|}.
\end{equation}
If we work with the (non-negative) rapidities $\zeta_i$ defined by
\begin{equation}
|v_i| = \tanh \zeta_i, 
\end{equation}
then
\begin{equation}
 \tanh\left| \zeta_1-\zeta_2 \right|  \leq |v_{12}| \leq  \tanh( \zeta_1 + \zeta_2).
\end{equation}
That is
\begin{equation}
 \tanh\left| \zeta_1-\zeta_2 \right|  \leq \tanh(\zeta_{12}) \leq  \tanh( \zeta_1 + \zeta_2).
\end{equation}
which implies
\begin{equation}
\left| \zeta_1-\zeta_2 \right|  \leq \zeta_{12} \leq   \zeta_1 + \zeta_2.
\end{equation}
It is this version that is closest in spirit to the Euclidean result, and this version that is more likely to lead to a suitable constraint on the composition of $n$ relative velocities.  
We could also write the 2-velocity constraint as
\begin{equation}
 \tanh\bigg| \tanh^{-1}|v_1|-\tanh^{-1}|v_2| \bigg|  \leq |v_{12}| \leq  \tanh\bigg( \tanh^{-1}|v_1| + \tanh^{-1}|v_2| \bigg).
\end{equation}

\clearpage
\subsection{Non-collinear velocities}

If the velocities are \emph{not collinear} there is a more complicated rule for combining velocities: 
\begin{equation}
\vec v_{12} = \vec v_1 \oplus \vec v_2. 
\end{equation}
Fortunately we will not need to be explicit about the details.
(For more details see for instance almost any medium-level technical book on special relativity~\cite{moller1952, jackson1998}, or for example references~\cite{O'Donnell:2011, stapp1956, fisher1972, ferraro1999}.) 
If we further define a rapidity \emph{vector} 
\begin{equation}
\vec \zeta = \left\{ \tanh^{-1}|v| \right\}  \; \hat v,
\end{equation}
there will be an analogous vectorial composition rule in rapidity space 
\begin{equation}
\vec\zeta_{12} = \vec\zeta_1 \boxplus \vec\zeta_2. 
\end{equation}
Fortunately we do not need the full power of the non-collinear composition rule, we only need to know the simple result obtained by looking at the extreme case of collinear (parallel/anti-parallel) motion:
\begin{equation}
\left| \;|\vec\zeta_1|- |\vec\zeta_2|\, \right|  \leq |\vec\zeta_1\boxplus\vec\zeta_2| \leq   |\vec\zeta_1| + |\vec\zeta_2|.
\end{equation}
That is:
\begin{equation}
\left| \;|\vec\zeta_1|- |\vec\zeta_2|\, \right|  \leq |\vec\zeta_{12}| \leq   |\vec\zeta_1| + |\vec\zeta_2|.
\end{equation}
So even for non-collinear motion we still have 
\begin{equation}
\left| \zeta_1-\zeta_2 \right|  \leq \zeta_{12} \leq   \zeta_1 + \zeta_2.
\end{equation}
We can now immediately apply the bound we have already derived for compound walks in physical Euclidean space.

\clearpage
\subsection{Bounds on the composition of velocities}
\subsubsection{Upper bounds}

For $n$ velocities the upper bound is straightforward, we just iterate the two-step result to obtain
\begin{equation}
\zeta_{12\cdots n} \leq   \sum_{i=1}^n\zeta_i, 
\end{equation}
whence
\begin{equation}
 |v_{12\cdots n}| \leq  \tanh\left[ \sum_{i=1}^n\zeta_i \right].
\end{equation}
We can also write this as
\begin{equation}
 |v_{12\cdots n}| \leq  \tanh\left[ \sum_{i=1}^n\tanh^{-1}|v_i| \right].
\end{equation}
Here are some explicit special cases obtained by straightforward manipulation of hyperbolic trig identities. Relativistically combining three velocities one has:
\begin{equation}
|v_{123}| \leq  {|v_1|+|v_2|+|v_3|+  |v_1||v_2||v_3|\over 1 + |v_1||v_2|+|v_2||v_3|+ |v_3||v_1|}.
\end{equation}
Similarly, relativistically combining four velocities one has:
\begin{equation}
|v_{1234}| \leq  {|v_1|+|v_2|+|v_3|+ |v_4| +  |v_1||v_2||v_3| +  |v_2||v_3||v_4| +  |v_3||v_4||v_1|+  |v_4||v_1||v_2|
\over 1 + |v_1||v_2|+|v_2||v_3|+ |v_3||v_4|  +|v_4||v_1| + |v_1||v_3|  + |v_2||v_4|   +   |v_1||v_2||v_3||v_4| } .
\end{equation}
If one additionally knows that all velocities are collinear, then instead of bounds one has the related \emph{equalities}
\begin{equation}
v_{123} =  {v_1+v_2+v_3+  v_1v_2v_3\over 1 + v_1v_2+v_2v_3+ v_3v_1},
\end{equation}
and
\begin{equation}
v_{1234} =   {v_1+v_2+v_3+ v_4 +  v_1v_2v_3 +  v_2v_3v_4 +  v_3v_4v_1+  v_4v_1v_2
\over 1 + v_1v_2+v_2v_3+ v_3v_4  +v_4v_1 + v_1v_3  + v_2v_4   +   v_1v_2v_3v_4 } .
\end{equation}
(There does not seem to be any more pleasant reformulation of these results, and in the completely general $n$-velocity case the general the ``tanh'' formula above seems to be the best one can do.)

\subsubsection{Lower bounds}

Obtaining an explicit lower bound is again a lot trickier than the upper bound. When relativistically combining three velocities then, (because of the monotonicity of the tanh function), one has
\begin{equation}
 |v_{123}| \geq  \tanh\left[ \vphantom{\bigg|}
 \max\left\{ \zeta_1 - \zeta_2-\zeta_3, \; \; \zeta_2 - \zeta_3-\zeta_1, \;\; \zeta_3 - \zeta_1-\zeta_2, \;\; 0 \right\}
 \right].
\end{equation}
When  relativistically combining $n$ velocities the best one can do is this:
\begin{equation}
 |v_{12\cdots n}| \geq  \tanh \left[\max\left\{ \zeta_i - \sum_{j \neq i }\zeta_j , \;\; 0 \right\}\right].
\end{equation}
We can also write this as
\begin{equation}
 |v_{12\cdots n}| \geq  \tanh \left[\max\left\{ 2 \zeta_i - \sum_{j=1}^n \zeta_j , \;\; 0 \right\}\right].
\end{equation}
Now defining
\begin{equation}
M_{12\cdots n} \equiv \tanh \left[\sum_{j=1}^n \zeta_j\right],
\end{equation}
and
\begin{equation}
v_\mathrm{peak} = \max_i \{ |v_i| \},
\end{equation}
and setting
\begin{equation}
m_{12\cdots n}  \equiv \tanh \left[\vphantom{\bigg|}\max\left\{ 2 \tanh^{-1} v_\mathrm{peak} - \tanh^{-1} M_{12\cdots n}   , \;\; 0 \right\}\right],
\end{equation}
we can also write this as
\begin{equation}
m_{12\cdots n}   \leq  |v_{12\cdots n}| \leq  M_{12\cdots n}.
\end{equation}
So there certainly are quite non-trivial constraints one can place on the relativistic combination of velocities, but they are a little less obvious than one might at first suspect.

\clearpage
\section{Scattering}

Compound scattering processes were extensively discussed in reference~\cite{Boonserm:2011}. (For additional background see~\cite{geometric, Sanchez-Soto, pedagogical, Peres, Kowalski, Korasani, SS1, SS2}; for various explicit bounds on transmission and reflection probabilities for scattering processes see references~\cite{Bounds0, Bounds-beta, Bounds-greybody, Bounds-mg, Bounds-analytic, Shabat-Zakharov, Bounds-thesis, Boonserm:2012, Boonserm:icast2012}; for a survey of exact results see reference~\cite{QNF:analytic}.) Rather than unnecessarily repeating the results of reference~\cite{Boonserm:2011}, we shall herein content ourselves with a few explicit comments regarding 2-barrier, 3-barrier, and 4-barrier systems. 
For two non-overlapping barriers the transmission and reflection probabilities are bounded by
\begin{equation}
 {T_1 T_2\over  \left\{ 1 + \sqrt{1-T_1}\sqrt{1-T_2} \right\}^2} \leq T_{12} \leq {T_1 T_2\over  \left\{ 1 - \sqrt{1-T_1}\sqrt{1-T_2} \right\}^2}; 
\end{equation}
and 
\begin{equation}
\left\{ {\sqrt{R_1} - \sqrt{R_2} \over 1 - \sqrt{R_1}\sqrt{R_2} }\right\}^2 \leq R_{12} \leq \left\{ {\sqrt{R_1} + \sqrt{R_2} \over 1 + \sqrt{R_1}\sqrt{R_2} }\right\}^2.
\end{equation}
For three non-overlapping barriers, the results of  reference~\cite{Boonserm:2011}, combined with a little work using hyperbolic trigonometric identities, lead to
\begin{equation}
T_{123} \geq {T_1 T_2 T_3 \over \left\{1 + \sqrt{(1-T_2)(1-T_3)}   + \sqrt{(1-T_3)(1-T_1)}   + \sqrt{(1-T_1)(1-T_2)}   \right\}^2 };
\end{equation}
and
\begin{equation}
R_{123} \leq \left\{ { \sqrt{R_1R_2R_3} + \sqrt{R_1} +\sqrt{R_2} + \sqrt{R_3}   \over 1 + \sqrt{R_2R_3}   + \sqrt{R_3R_1}   + \sqrt{R_1R_2} }  \right\}^2.
\end{equation}
For four non-overlapping barriers, a completely analogous calculation straightforwardly yields
\begin{equation}
T_{1234} \geq {T_1 T_2 T_3 T_4\over \left\{1 + \sum_{i<j}\sqrt{(1-T_i)(1-T_j)}   + \sqrt{(1-T_1)(1-T_2)(1-T_3)(1-T_4)}    \right\}^2 };
\end{equation}
and 
\begin{equation}
R_{1234} \leq \left\{ { \sqrt{R_1} +\sqrt{R_2R_3R_4}   + \hbox{(cyclic permutations)}  \over 1 + \sum_{i<j}\sqrt{R_iR_j}  + \sqrt{R_1R_2R_3R_4} }  \right\}^2.
\end{equation}
That is, explicitly,
\begin{eqnarray}
&&\hspace{-20pt}
R_{1234} \leq\hfill
\\
&&
\hspace{-20pt}
\left\{ { \sqrt{R_1}+ \sqrt{R_2}+ \sqrt{R_3}+ \sqrt{R_4} 
 +  \sqrt{R_2R_3R_4} + \sqrt{R_3R_4R_1} + \sqrt{R_4R_1R_2}  +\sqrt{R_1R_2R_3}
  \over 1 + \sum_{i<j}\sqrt{R_iR_j}  + \sqrt{R_1R_2R_3R_4} }  \right\}^2.
\nonumber
\end{eqnarray}
Upper bounds on $T$, and lower bounds on $R$, are less algebraically tractable, (at least in explicit closed form), and we refer the reader to reference~\cite{Boonserm:2011} for more details. 

\section{Parametric excitations}

By working in the temporal rather than spatial domain, particle scattering processes can be re-phrased in terms of particle production via parametric excitation. (See reference~\cite{Boonserm:2011} for details). In this context, the net particle production due to two non-overlapping excitation events is bounded by
\begin{equation}
\left\{\sqrt{N_1(N_2+1)}-   \sqrt{N_2(N_1+1)}\right\}^2 \leq N_{12} \leq \left\{\sqrt{N_1(N_2+1)}+   \sqrt{N_2(N_1+1)}\right\}^2.
\end{equation}
For three non-overlapping excitation events one obtains
\begin{eqnarray}
N_{123} &\leq& \Big\{ \sqrt{N_1(1+N_2)(1+N_3)} + \sqrt{N_2(1+N_3)(1+N_1)} 
\nonumber\\
&&\qquad
+ \sqrt{N_3(1+N_1)(1+N_2)} + \sqrt{N_1N_2N_3}  \Big\}^2.
\end{eqnarray}
For four non-overlapping excitation events a straightforward (but rather tedious) calculation yields
\begin{eqnarray}
N_{1234} &\leq& \Big\{ \sqrt{N_1(1+N_2)(1+N_3)(1+N_4)}  + \sqrt{N_1N_2N_3(1+N_4)}  
\nonumber\\
&& \qquad
+ \hbox{ (cyclic permutations)} \Big\}^2.
\end{eqnarray}
Further ``explicit'' algebraic formulae would be rather unwieldy, and for all practical purposes one is better off using the somewhat less ``explicit'' formulae in presented terms of hyperbolic functions in reference~\cite{Boonserm:2011}. Similarly lower bounds on $N$ are less algebraically tractable, (at least in explicit closed form), and we again refer the reader to reference~\cite{Boonserm:2011} for more details.

\section{Discussion}

That particle scattering in the spatial domain is mathematically  intimately related to particle production in the temporal domain is a very standard result, ultimately going back to the relationship between scattering and transmission amplitudes and the Bogoliubov coefficients.  (See for instance references~\cite{Boonserm:2011, geometric, pedagogical, QNF:analytic} for more details on this specific point.) The intimate mathematical relationship between particle scattering and relativistic composition of velocities is less well-known, but is quite standard.  The $SO(3,1)$ Lorentz group is locally isomorphic to $SL(2,\C)$, while  the group of transfer matrices $SU(1,1)$ is locally isomorphic to $SL(2,\R)$.  Ultimately it is the fact that  their Lie algebras are both isomorphic to Euclidean space that ties the three problems (physical Euclidean space, relativistic composition of velocities, and composition of scattering processes) together.
The overall result of the current article is to rigorously establish several clearly motivated and robust mathematical bounds on these three closely inter-related physical problems.


\section*{Acknowledgments}
MV was supported by the Marsden Fund, and by a James Cook fellowship, both administered by the Royal Society of New Zealand. PB was supported by a scholarship from the Royal Government of Thailand, and partially supported by a travel grant from FQXi, and by a grant for the professional development of new academic staff from the Ratchadapisek Somphot Fund at Chulalongkorn University, by the Thailand Toray Science Foundation (TTSF), by the Thailand Research Fund (TRF), the Office of the Higher Education Commission (OHEC), Chulalongkorn University (MRG5680171), and by the Research Strategic plan program (A1B1), Faculty of Science, Chulalongkorn University. 


\end{document}